# The FKPP wave front as a sensor of perturbed diffusion in concentrated systems

Gabriel Morgado[1,2], Bogdan Nowakowski[1], and Annie Lemarchand[2]

[1]Institute of Physical Chemistry, Polish Academy of Sciences, Kasprzaka 44/52, 01-224 Warsaw, Poland
[2]Sorbonne Université, CNRS UMR 7600, Laboratoire de Physique Théorique de la Matière Condensée, 4 place Jussieu, case courrier 121, 75252 Paris CEDEX 05, France

November 13, 2018

Corresponding author: Annie Lemarchand, E-mail: anle@lptmc.jussieu.fr

### Abstract

The sensitivity to perturbations of the Fisher and Kolmogorov, Petrovskii, Piskunov front is used to find a quantity revealing perturbations of diffusion in a concentrated solution of two chemical species with different diffusivities. The deterministic dynamics includes cross-diffusion terms due to the deviation from the dilution limit. The behaviors of the front speed, the shift between the concentration profiles of the two species, and the width of the reactive zone are investigated, both analytically and numerically. The shift between the two profiles turns out to be a well-adapted criterion presenting noticeable variations with the deviation from the dilution limit in a wide range of parameter values.

# 1 Introduction

The Fisher and Kolmogorov, Petrovskii, Piskunov (FKPP) wave front is the prototype of pulled fronts, whose properties are strongly influenced by the leading edge of the profile [1, 2]. From the perspective of applications, wave fronts of FKPP type are widely used in models of combustion [3] and biology [4, 5, 6], in particular to account for adaptation, mutation and selection in evolutionary strategies. Sufficiently steep initial profiles converge to the front propagating at the minimum velocity deduced from a linear stability analysis [7]. The FKPP front is known to be highly sensitive to even small perturbations of many different origins. Brunet and Derrida proved that a small cutoff introduced in the leading edge of the front induces a negative correction to the front speed [8]. The description of a reaction-diffusion system at a mesoscopic scale by a master equation [12, 13] as well as particle dynamics simulations using the Direct Simulation Monte Carlo (DSMC) method [14] both revealed that the discrete nature of particle numbers induces analogous corrections to the front speed as a cutoff in the deterministic partial differential equations. Roughly speaking, the cutoff can be interpreted as the inverse of the particle number in the reactive zone [15]. It has then been shown in the framework of a Langevin approach that the effect of a multiplicative noise on the front speed can be studied under the scope of a modified cutoff theory [10, 11]. The effect of a slightly exothermic reaction on the front speed has also been studied using DSMC. Below a critical heat release, the speed remains equal to the one in the isothermal case and is imposed by the Chapman-Jouguet criterion above it [16]. In addition, the DSMC method has been used to study the impact of the perturbation of local equilibrium by a fast reaction associated with a small activation energy: Reaction-induced non Maxwellian particle velocity distributions result in positive corrections to the front speed [17]. Molecular dynamics simulations of dense fluids also lead to propagation speeds larger than the marginally stable one [15]. Recent articles focus on the effect of an advection term [18, 19, 20].

In this paper, we study a reaction-diffusion wave front of FKPP type involving two chemical species A and B of different diffusivities. At high concentrations, the partial dif-ferential equations involve cross-diffusion terms according to linear nonequilibrium thermodynamics [21, 22, 23]. Cross-diffusion phenomena have been experimentally characterized in microemulsions [24, 25, 26]. They are known to induce hydrodynamic instabilities in reaction-diffusion-convection patterns in microemulsion [27, 28]. We recently checked that the wavelength of a Turing pattern is not affected and can therefore not characterize the perturbation of diffusion induced by high concentrations [23, 29]. The goal of the paper is to harness the sensitivity of FKPP fronts to find a macroscopic quantity depending on the detail of the diffusion rates and thus sensitive to the deviation from the dilution limit. Literature mostly reports on corrections to the front speed [9, 10, 11, 18, 19, 20, 30]. We will first examine the impact of diffusion perturbation on the front speed and then investigate the behavior of alternative quantities.

The paper is organized as follows. We present the reaction-diffusion model in Sec. 2. Analytical expressions of different quantities characterizing the wave front in dilute and concentrated systems are derived in Sec. 3. Specifically, we look for effects of diffusion perturbation on the propagation speed, the shift between A and B profiles, and the width of the reactive zone. The analytical predictions are compared to numerical results in Sec. 4. Section 5 contains conclusions.

## 2 The reaction-diffusion model

The system is composed of three chemical species. Two of them, A and B, are reactive whereas the third species S is the solvent. The reaction scheme is given by

$$\text{A} + \text{B} \xrightarrow{k} 2\,\text{A} \tag{1}$$

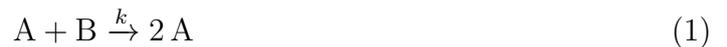

where $k$ is a rate constant. In a dilute system, the reaction-diffusion equations associated with the mechanism are

$$\partial_t A = kAB + D_A \partial_x^2 A \tag{2}$$

$$\partial_t B = -kAB + D_B \partial_x^2 B \tag{3}$$

where $A$ and $B$ are the concentrations of species A and B, and $D_A$ and $D_B$ are the diffusion coefficients of A and B species, respectively. In a concentrated solution, diffusion of A and B species may be perturbed. Nevertheless, we assume that the solution is ideal, in the sense that the activity remains equal to the concentration. In the framework of linear nonequilibrium thermodynamics, we derived linear relations between generalized diffusion fluxes and forces which couple the diffusion of a species with the gradients of each constituent of the mixture [21, 23]. After the elimination of the solvent concentration $S$, we have shown that the reaction-diffusion equations are given by:

$$\partial_t A = kAB + D_A \partial_x \left[\left(1 - \frac{A}{C}\right)\partial_x A\right] - D_B \partial_x \left(\frac{A}{C}\partial_x B\right) \quad (4)$$

$$\partial_t B = -kAB + D_B \partial_x \left[\left(1 - \frac{B}{C}\right)\partial_x B\right] - D_A \partial_x \left(\frac{B}{C}\partial_x A\right) \quad (5)$$

where the total concentration $C = A + B + S$ is constant. Equations (4) and (5) converge to Eqs. (2) and (3) in the dilution limit $(A+B)/C \to 0$ and are valid for sufficiently small values of $(A+B)/C$.

We choose inhomogeneous initial conditions in the form of a step function for species A and B:

$$\begin{cases} x < 0, & A(x, t=0) = V_0, \quad B(x, t=0) = 0 \\ x \geq 0, & A(x, t=0) = 0, \quad B(x, t=0) = V_0 \end{cases} \quad (6)$$

where the constant $V_0$ characterizes the height of the step. The reaction-diffusion equations have wave front solutions which propagate at constant speed $v^\alpha$ where the exponent $\alpha = d$ for the dilute system and $\alpha = c$ for the concentrated system. These FKPP fronts are also called pulled fronts because the speed is determined by the leading edge of the profile which pulls the bulk to the right [7, 8].

For identical diffusion coefficients $D_A = D_B$ and initially homogeneous conditions for $S$ and $A + B = V_0$, the sum $A + B$ does not evolve. Then, introducing the conservation relation $A + B = V_0$ into Eqs. (4) and (5), we find that the concentrated solution obeys the same unperturbed equations given in Eqs. (2) and (3) as the diluted system. With the aim of specifying how the properties of a FKPP wave front are perturbed as the system becomes more concentrated, we consider different diffusion coefficients $D_A$ and $D_B$ in the following.

# 3 Analytical derivation of wave front features in dilute and concentrated systems

## 3.1 Propagation speed

To derive an approximate analytical expression of the propagation speed, we perform a linear stability analysis around the steady state $(A = 0, B = V_0)$ in the moving frame at speed $v^\alpha$. Linearizing the reaction-diffusion equations is supposed to be valid in the leading edge of the front. We introduce the following transformation:

$$\xi = \frac{x}{v^\alpha} - t \tag{7}$$

$$A(x,t) = f^\alpha(\xi) \tag{8}$$

$$B(x,t) = g^\alpha(\xi) \tag{9}$$

where $\alpha = d, c$.

We first address the case of a dilute system. Equations (2) and (3) can be rewritten as

$$0 = kf^d g^d + (f^d)' + \varepsilon^d D_A (f^d)'' \tag{10}$$

$$0 = -kf^d g^d + (g^d)' + \varepsilon^d D_B (g^d)'' \tag{11}$$

where $\varepsilon^d = 1/v^{d2}$ and $'$ denotes the derivation with respect to $\xi$. The second-order differential equations are transformed into first-order equations in the four-dimension space $(f^d, (f^d)', g^d, (g^d)')$. We perform a linear stability analysis around the unstable steady state and obtain the following linearized uncoupled system for $(f^d, (f^d)')$:

$$\frac{df^d}{d\xi} = (f^d)' \tag{12}$$

$$\frac{d(f^d)'}{d\xi} = -\frac{v^{d2}}{D_A}\left(kV_0 f^d + (f^d)'\right) \tag{13}$$

which leads to the eigenvalues $\lambda_\pm$:

$$\lambda_\pm = \frac{-v^{d2} \pm v^d \sqrt{v^{d2} - 4kV_0 D_A}}{2D_A} \tag{14}$$

The existence of wave front solutions is ensured for real eigenvalues which imposes the minimum velocity:

$$v^d_{\min} = 2\sqrt{kV_0 D_A} \tag{15}$$

In the concentrated case, Eqs. (4) and (5) read

$$0 = kf^c g^c + (f^c)' + \varepsilon^c \left( D_A \left[ \left(1 - \frac{(f^c)'}{C}\right)(f^c)'' - \frac{(f^c)'^2}{C} \right] - D_B \left[ \frac{f^c (g^c)''}{C} + \frac{(f^c)'(g^c)'}{C} \right] \right) \quad (16)$$

$$0 = -kf^c g^c + (g^c)' + \varepsilon^c \left( D_B \left[ \left(1 - \frac{g^c}{C}\right)(g^c)' - \frac{(g^c)'^2}{C} \right] - D_A \left[ \frac{g^c (f^c)''}{C} + \frac{(f^c)'(g^c)'}{C} \right] \right) \quad (17)$$

where $\varepsilon^c = 1/v^{c2}$. Following the same procedure as in the dilute case, we find

$$v^c_{\min} = 2\sqrt{kV_0 D_A} \quad (18)$$

Hence, the front speeds in the dilute system and the concentrated system are identical in the framework of a linear stability analysis. Consequently, the following notations are introduced:

$$v^c_{\min} = v^d_{\min} = v \quad (19)$$

$$\varepsilon^c = \varepsilon^d = \varepsilon \quad (20)$$

We checked that for sufficiently steep initial conditions and after a transient regime, the wave front propagates at the minimum speed $v$, as in the case of identical diffusion coefficients [7, 8]. Interestingly, in both the dilute and concentrated systems, the minimum propagation speed of the linearized system does not depend on the diffusion coefficient $D_B$ of species B and only depend on the product $kV_0 D_A$.

## 3.2 Front profile

A perturbation technique is used to determine analytical expressions of quantities characterizing the wave front profile. We look for solutions of the reaction-diffusion equations in the frame moving at front speed $v$ as a Taylor expansion in the small parameter $\varepsilon$ [4]. As $\varepsilon$ tends to zero, Eqs. (10) and (11) and Eqs. (16) and (17) switch from second-order differential equations to first-order equations. The boundary conditions of the first-order equations must be compatible with the ones of the second-order equations. However, the reactive terms $\pm k f^\alpha g^\alpha$ and the first-order terms $f^{\alpha\prime}$ and $g^{\alpha\prime}$ equal zero at the boundaries $\xi = \pm\infty$ for all perturbation orders, which ensures the consistency of a regular

perturbation procedure:

$$f^\alpha = f_0^\alpha + \varepsilon f_1^\alpha + \varepsilon^2 f_2^\alpha + ... \tag{21}$$

$$g^\alpha = g_0^\alpha + \varepsilon g_1^\alpha + \varepsilon^2 g_2^\alpha + ... \tag{22}$$

where $f_i^\alpha$ and $g_i^\alpha$ are the $i$-th order corrections with $i = 0, 1, 2, ...$ and $\alpha = d, c$. The boundary conditions obey:

$$f_0^\alpha(-\infty) = V_0, \quad f_0^\alpha(+\infty) = 0 \tag{23}$$

$$g_0^\alpha(-\infty) = 0, \quad g_0^\alpha(+\infty) = V_0 \tag{24}$$

$$f_i^\alpha(\pm\infty) = g_i^\alpha(\pm\infty) = 0, \quad \text{for } i \geq 1 \tag{25}$$

The origin of the $\xi$-axis is chosen such that

$$f_0^\alpha(0) = \frac{V_0}{2} \tag{26}$$

$$f_i^\alpha(0) = 0, \quad \text{for } i \geq 1 \tag{27}$$

The zeroth-order solutions are straightforwardly deduced from Eqs. (10) and (11) and Eqs. (16) and (17) without diffusion terms

$$f_0^\alpha = \frac{V_0}{1 + e^{kV_0\xi}} \tag{28}$$

$$g_0^\alpha = \frac{V_0}{1 + e^{-kV_0\xi}} \tag{29}$$

for $\alpha = d, c$.

Instead of determining the higher-order solutions, we focus on characteristic properties of the profiles. We define the height $h^\alpha = f^\alpha(0) - g^\alpha(0)$ as the difference of concentrations between species A and B in the moving frame at the origin $\xi = 0$. The height $h^\alpha$ evaluates the shift between the profiles of species A and B due to their different diffusivities. Using Eqs. (26) and (27), we obtain the evaluations of the height up to the first and second orders:

$$h_1^\alpha = \varepsilon g_1^\alpha(0) \tag{30}$$

$$h_2^\alpha = \varepsilon g_1^\alpha(0) + \varepsilon^2 g_2^\alpha(0) \tag{31}$$

for $\alpha = d, c$. Using Eqs. (10) and (11), we obtain the first $h_1^d$ and second-order $h_2^d$ approximations of the height in the dilute case:

$$h_1^d = \frac{V_0}{16}\left(1 - \frac{D_B}{D_A}\right) \tag{32}$$

$$h_2^d = \frac{V_0}{16}\left(1 - \frac{D_B}{D_A}\right)\left[1 + \frac{1}{8}\left(1 - \frac{D_B}{D_A}\right)\right] \tag{33}$$

As a result, the height $h_2^d$ does not depend on the rate constant $k$ and the scaled height $h_2^d/V_0$ only depends on the ratio $D_B/D_A$. Using Eqs. (16) and (17), we find in the concentrated case:

$$h_1^c = \frac{V_0}{16}\left(1 - \frac{D_B}{D_A}\right)\left(1 - \frac{V_0}{C}\right) \tag{34}$$

$$h_2^c = \frac{V_0}{16}\left(1 - \frac{D_B}{D_A}\right)\left(1 - \frac{V_0}{C}\right)\left[1 + \frac{1}{8}\left(1 - \frac{D_B}{D_A}\right)\left(1 - 2\frac{V_0}{C}\right)\right] \tag{35}$$

We check that, in the dilution limit $\frac{V_0}{C} \to 0$, the first and second-order expressions of the height in the concentrated system given in Eqs. (34) and (35) converge to the first and second-order expressions of the height in the dilute system given in Eqs. (32) and (33). Up to the second order, the scaled height $h^c/V_0$ only depends on the ratio of the diffusion coefficients $D_B/D_A$ and the deviation $V_0/C$ from the dilution limit. The parameter $V_0/C$ varying in the range $[0, 1]$, the first-order evaluation $h_1^c$ in a concentrated system is always smaller than the corresponding quantity $h_1^d$ given in Eq. (32) in the dilute system. High concentrations tend to reduce the shift between A and B profiles, at least at the first order.

The width $W^\alpha$ of the wave front is deduced from the inverse of the steepness of the A profile at $\xi = 0$

$$W^\alpha = -vV_0/(f^\alpha)'(0) \tag{36}$$

We consider the following evaluations of the width:

$$W_1^\alpha = \frac{-vV_0}{(f_0^\alpha)'(0) + \varepsilon(f_1^\alpha)'(0)} \tag{37}$$

$$W_2^\alpha = \frac{-vV_0}{(f_0^\alpha)'(0) + \varepsilon(f_1^\alpha)'(0) + \varepsilon^2(f_2^\alpha)'(0)} \tag{38}$$

deduced from the first-order and the second-order expansions of $f^\alpha$. For the sake of simplicity, $W_1^\alpha$ and $W_2^\alpha$ will be called first and second-order evaluations of the width, respectively. In the dilute case, the first-order $W_1^d$ and the second-order $W_2^d$ expressions of the width are deduced from Eqs. (10) and (11):

$$W_1^d = 8\sqrt{\frac{D_A}{kV_0}}\left[1+\frac{1}{8}\left(1-\frac{D_B}{D_A}\right)\right]^{-1} \tag{39}$$

$$W_2^d = 8\sqrt{\frac{D_A}{kV_0}}\left[1+\frac{1}{8}\left(1-\frac{D_B}{D_A}\right)-\frac{1}{64}\frac{D_B}{D_A}\left(3-\frac{D_B}{D_A}\right)\right]^{-1} \tag{40}$$

It is worth noting that $8\sqrt{\frac{D_A}{kV_0}}$ is an approximation of the width in the case $D_A = D_B$ known to be valid up to the first order [4]. The second-order evaluation $W_2^d$ provides the corrected expression $8\sqrt{\frac{D_A}{kV_0}}\left(1-\frac{1}{32}\right)$ of the width for $D_A = D_B$. Using Eqs. (16) and (17) in the concentrated case, we obtain after tedious calculations:

$$W_1^c = 8\sqrt{\frac{D_A}{kV_0}}\left[1+\frac{1}{8}\left(1-\frac{D_B}{D_A}\right)\left(1-\frac{3V_0}{2C}\right)\right]^{-1} \tag{41}$$

$$W_2^c = 8\sqrt{\frac{D_A}{kV_0}}\left[1+\frac{1}{8}\left(1-\frac{D_B}{D_A}\right)\left(1-\frac{3V_0}{2C}\right)\right.$$
$$\left.-\frac{1}{64}\left[\frac{D_B}{D_A}\left(3-\frac{D_B}{D_A}\right)+\left(\frac{9}{2}-8\frac{D_B}{D_A}+\frac{7}{2}\frac{D_B^2}{D_A^2}\right)\frac{V_0}{C}-\left(\frac{7}{2}-7\frac{D_B}{D_A}+\frac{7}{2}\frac{D_B^2}{D_A^2}\right)\frac{V_0^2}{C^2}\right]\right]^{-1} \tag{42}$$

We check that, in the dilution limit $\frac{V_0}{C} \to 0$, the first and second-order expressions of the width in the concentrated system given in Eqs. (41) and (42) converge to the first and second-order expressions of the width in the dilute system given in Eqs. (39) and (40). The scaled width $\sqrt{\frac{kV_0}{D_A}}W^c$ only depends on $D_B/D_A$ and $V_0/C$ as the height does. In the next section, the analytical predictions of the height and the width are compared to the corresponding numerical results deduced from Eqs. (2) and (3) in the dilute case and Eqs. (4) and (5) in the concentrated case.

## 4 Comparison between analytical and numerical results

Analytical results are derived from expansions with respect to $\varepsilon = \frac{1}{v^2}$. For the domain of validity of approximations to be the same for all the considered parameter values, we impose that the front speed remains constant, i.e. $k$, $V_0$, and $D_A$ are constant. In addition, the values of $k$, $V_0$, and $D_A$ are set such that $\varepsilon$ is much smaller than 1:

$$k = 10, \quad V_0 = 10, \quad D_A = 1 \tag{43}$$

According to Brunet and Derrida [8], a small cutoff $\delta$ introduced in the nonlinear reactive term $\pm kAB$ induces a negative correction to the propagation speed:

$$\frac{v - v_s}{v} = \frac{\pi^2}{2(\ln \delta)^2} \qquad (44)$$

where $v_s$ is the velocity of the simulated front. More generally, FKPP wave fronts are known to be sensitive to small perturbations, including fluctuations [15] in stochastic descriptions, perturbation of velocity distribution function [17] in particle dynamics simulations. With the aim of unambiguously assigning the observed perturbations of a wave front to high concentrations, the effect of a cutoff on the numerical results has to be evaluated and disregarded. If sufficiently fine space and time discretizations are employed, the cutoff mainly originates from the precision of the computations involving real numbers. Choosing double precision leads to a cutoff $\delta \simeq 10^{-16}$. Using Eq. (44), the relative correction to the front speed is estimated at 0.4%. The effect of space discretization is similar on wave fronts with profile widths occupying the same number of spatial cells. According to Eq. (39) and (41), at zeroth order, the width $W_0^\alpha = 8\sqrt{\frac{D_A}{kV_0}}$ only depends on the rate constant $k$, the diffusion coefficient $D_A$ of species A, and the boundary condition $V_0$. For the effect of the cutoff to be the same in all the numerical solutions for different parameters, we impose the cell length, $\Delta x = \frac{\pi W_0^\alpha}{5000}$, the total number of cells, $n = 50000$, and the time step, $\Delta t = \frac{0.1 \Delta x^2}{D_B^{\max}}$, where $D_B^{\max} = 16$ is the maximum diffusion coefficient considered. Hence, the width occupies about 1600 cells in all cases. For the initial condition defined in Eqs. (6), we numerically solve Eqs. (2) and (3) and Eqs. (4) and (5) using Euler method for different values of the diffusion coefficient $D_B$ in the interval $\left[\frac{1}{16}, 16\right]$ and the total concentration $C$ in the interval $[25, 400]$.

To mimic an infinite system in the $x$-direction, it is necessary to counterbalance the production of species A due to the propagation of the reactive front. At each time step where the sum of the concentrations of species A in each cell reaches the initial value $nV_0/2$, the first left cell is suppressed and a new cell is added to the right with no A species and a $V_0$ concentration for species B. This trick amounts to switching in the moving frame at the propagation speed of the wave front. The speed is numerically evaluated using the time difference between two creations of right cells after the stationary regime has been reached.

For these conditions, we find that the front speeds associated with the dilute and the concentrated cases are the same and about 0.5% smaller than the zeroth-order prediction $v_{\min}$. This result is close to the estimation of the cutoff effect induced by double precision. The choice of the other parameters, such as cell length and time step, and the simulation procedure are therefore satisfying. In addition, the numerical results confirm that the propagation speed is not impacted by the perturbation of diffusion induced by high concentrations as predicted in Eq. (18).

Figure 1 shows the stationary concentration profiles of A and B species deduced from the numerical integration of Eqs. (2) and (3) in the dilute case. In agreement with Eq. (33), a large ratio $D_B/D_A$ is chosen for the height $h^d$ illustrating the shift between A and B profiles to be sufficiently large. Whereas the A concentration abruptly vanishes in the leading edge, the B concentration smoothly tends to $V_0$. A and B profiles are noticeably asymmetric with respect to the $A = \dfrac{V_0}{2}$ axis. This feature disappears for $D_A = D_B$.

Figure 2a shows the variation of the scaled height $h^d/V_0$ in the dilute case with respect to the ratio of the diffusion coefficients $D_B/D_A$ in logarithmic scale. The first and second-order analytical expressions $h_1^d$ and $h_2^d$ given in Eq. (32) and (33) are compared to the results deduced from the numerical integration of Eqs. (2) and (3). The uncertainty on the numerical results due to discretization is smaller than the symbols.

As expected, the height vanishes for $D_B = D_A$. For $D_B/D_A < 1$, the height $h^d$ is positive and the first-order expression already offers a satisfying approximation. Considering that $D_A$ is set at 1 whereas $D_B$ varies, and that the perturbative term in Eq. (11) is proportional to $\varepsilon^d D_B = \dfrac{1}{4kV_0}\dfrac{D_B}{D_A}$, the analytical result is correct provided that $D_B$ is smaller or equal to 1, i.e. $D_B < D_A$. It is worth noting that, as $D_B \to 0$, the height $h^d$ tends to a positive limit slightly larger than $V_0/16$ as predicted by Eq. (33): For fixed B particles, the concentration of species B in the moving frame reaches the value $h^d \simeq V_0/16$, independent of the diffusion coefficient $D_A$ of species A at the abscissa $\xi = 0$ for which the concentration of species A equals $V_0/2$.

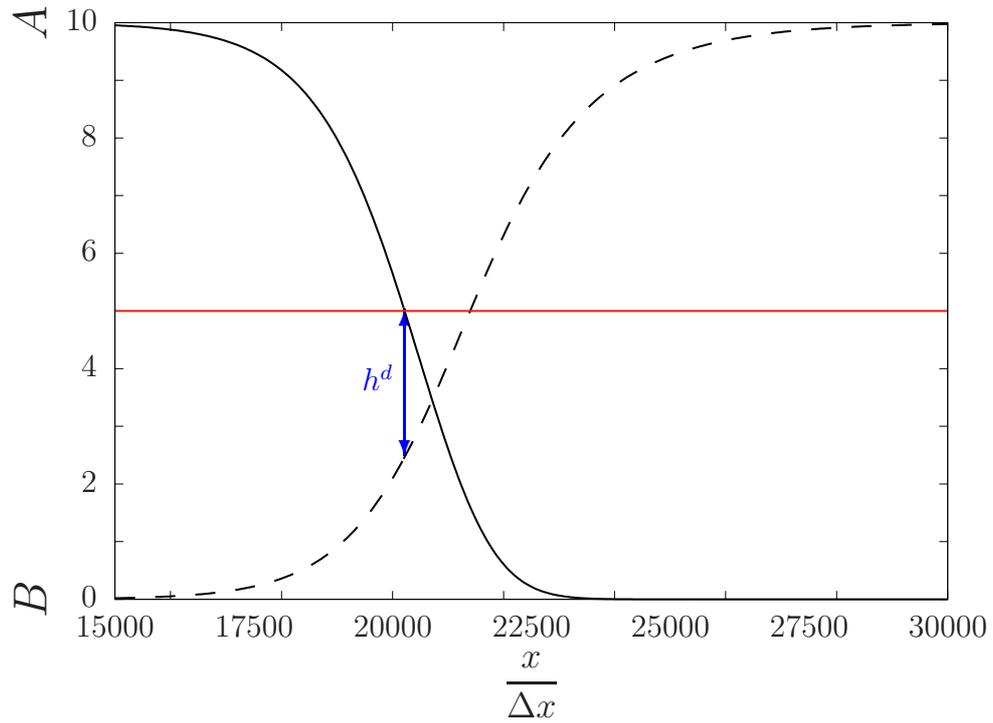

Figure 1: Snapshot of the wave front profile deduced from Eqs. (2,3) with $D_B/D_A = 16$ at time $t = 400$. Concentration profiles of A (black solid line) and B (black dashed line) versus spatial coordinate $\dfrac{x}{\Delta x}$. The horizontal line represents $\dfrac{V_0}{2}$. The vertical segment represents the height $h^d$ at $\xi = 0$.

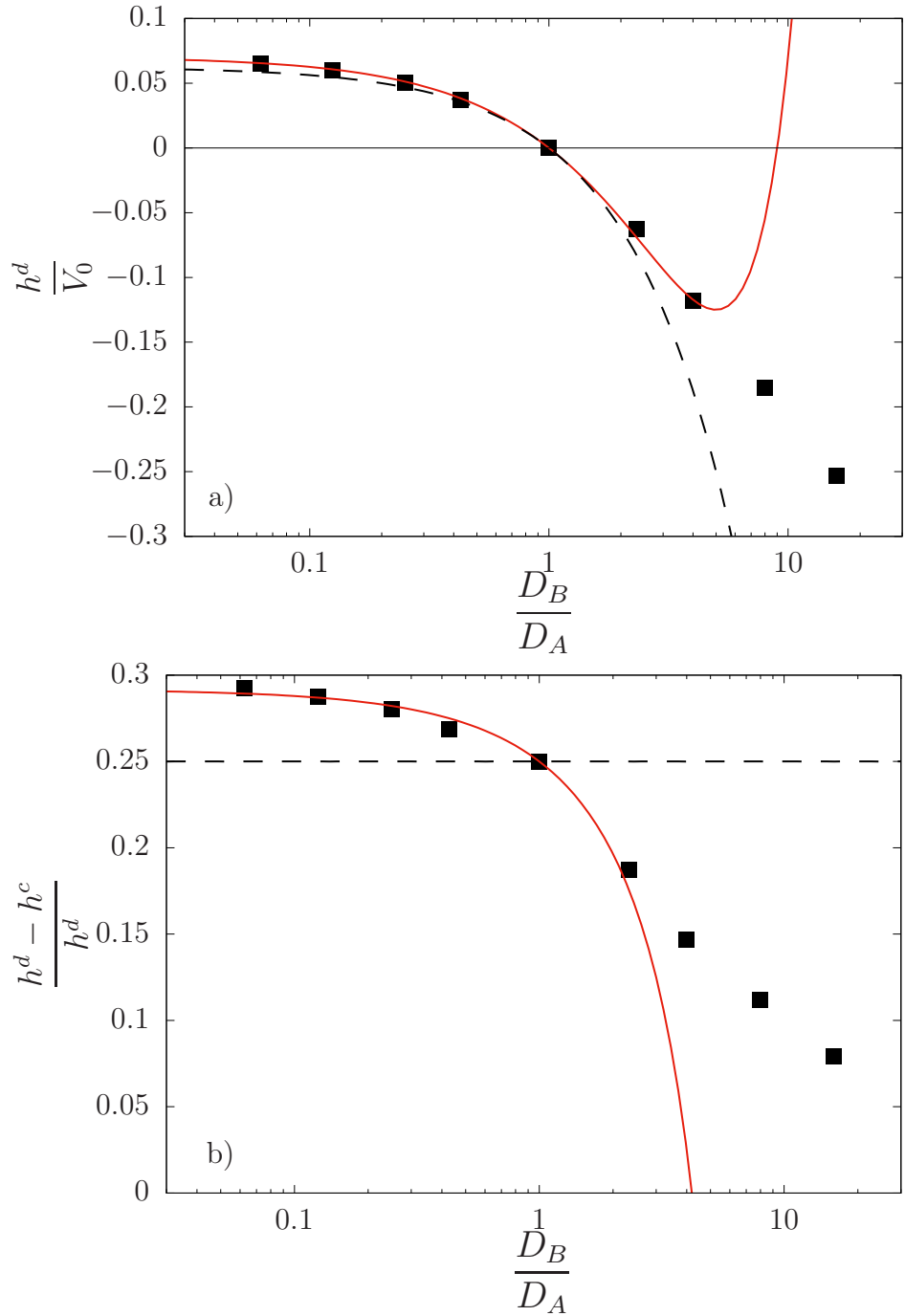

Figure 2: (a) Scaled height $h^d/V_0$ in the dilution limit and (b) relative difference between the dilute case and the concentrated case for $V_0/C = 0.25$ versus diffusion coefficients ratio $D_B/D_A$. First-order approximation (black dashed line), second-order approximation (red solid line), and numerical results (black squares).

For $D_B/D_A > 1$, the height $h^d$ is negative. The second-order approximation is valid until $D_B/D_A = 4$ and diverges for larger values of $D_B/D_A$. In the explored range of $D_B/D_A$, the height $h^d$ significantly decreases. Still, the scaled height $h^d/V_0$ is bounded by $-0.5$ since concentrations cannot be negative.

The behavior of the height $h^c$ associated with the concentrated system is similar to the one of $h^d$. Nevertheless, as shown in Fig. 2b, the relative height difference $\dfrac{h^d - h^c}{h^d}$ is always positive in the entire range of $D_B/D_A$ values. Hence, the shift between the concentration profiles of species A and B induced by the difference between the diffusion coefficients $D_A$ and $D_B$ is reduced in a concentrated system. Actually, according to Eqs. (4) and (5), the diffusion of a given species is affected by the diffusion coefficient of the other species which reduces the effect due to $D_A \neq D_B$. In the limit of large $D_B$, both $h^d$ and $h^c$ reach the extreme value $-0.5$ so that the difference $h^d - h^c$ tends to zero. As $D_B \to 0$, the relative height difference $\dfrac{h^d - h^c}{h^d}$ tends to a positive limit larger than the prediction $V_0/C$ of the first-order approximation deduced from Eqs. (32) and (34). The first-order approximation does not account for the variation of the relative height difference $\dfrac{h^d - h^c}{h^d}$ with respect to $D_B/D_A$. The numerical results perfectly agree with the second-order prediction in the domain $D_B/D_A < 1$ for which perturbative analysis is valid. Interestingly, the relative difference of heights reaches 28% for small $D_B/D_A$ values, making the shift between A and B profiles well adapted to the discrimination between the concentrated and the dilute system.

As mentioned in Sec. 3, the height $h^c$ only depends on $D_B/D_A$ and $V_0/C$. The variations of the height $h^c$ in the concentrated system with respect to $V_0/C$ are given in Fig. 3 for two different values of $D_B/D_A$. The parameter $V_0/C$ quantifies the deviation from the dilution limit obtained for $V_0/C \to 0$. We recall that a concentrated system does not refer to large values of $C$ but to high concentrations $A + B$ of the solute. The results shown in Fig. 2b are given for $V_0/C = 0.25$ which is a good compromise between a too small value for which concentration effects would be negligible and a too large value for which the reaction-diffusion equations (Eqs. (4) and (5)) would not be valid. In Fig. 3a, the results are given for $D_B/D_A = 1/16 < 1$, associated with a positive value of $h^c$ in agreement with the results shown in Fig. 2. We find that the shift between the profiles of A and B species decreases as the system becomes more concentrated, i.e. as $V_0/C$ increases. For the small value of $D_B/D_A$ chosen, the expansion technique rapidly converges and the agreement between the numerical results and the second-order approximation is excellent. In Fig. 3b, for $D_B/D_A = 7/3 > 1$, the height $h^c$ is negative and decreases in absolute value as $V_0/C$ increases. As already mentioned, the perturbation analysis is less relevant and the second-order prediction deviates from the numerical results. The concentrated system is closer to the standard FKPP front with $D_A = D_B$ than the dilute system is.

Figure 4a gives the variation of the front width $W^d$ of species A with respect to the ratio of the diffusion coefficients $D_B/D_A$ in logarithmic scale in the dilute case. The slope $s^d$ of the A concentration profile at $\xi = 0$, deduced from the numerical integration of Eqs. (2) and (3), is evaluated according to:

$$s^d = -\frac{0.2V_0}{\xi_2 - \xi_1} \qquad (45)$$

where the abscissa $\xi_1$ and $\xi_2$ in the moving frame obey $f^d(\xi_1) = 0.6V_0$ and $f^d(\xi_2) = 0.4V_0$. Equation (36) is used to obtain an estimation of the width of the reactive front. Both the first and second-order approximations given in Eqs. (39) and (40) satisfactorily agree with the numerical results for $D_B/D_A < 4$.

An analogous procedure is followed to determine an estimation of the width $W^c$ in the concentrated system from the numerical integration of Eqs. (4) and (5). The relative difference $(W^d - W^c)/W^d$ between the widths in the dilute system and the concentrated system versus $D_B/D_A$ is given in Fig. 4b. It is worth noting that it has been necessary to decrease cell length $\Delta x$ for $W^c \Delta x$ to reach about 1600 in order to obtain a sufficient precision in the numerical estimation of the relative correction. This constraint has motivated the choice of the cell length used in this study. For $D_B/D_A \leq 1$, a satisfying agreement is obtained between the numerical results and the first and second-order analytical predictions given in Eqs. (39) and (40) in the dilute case and Eqs. (41) and (42) in the concentrated case. Interestingly, the relative correction to the width induced by high concentrations changes sign as $D_B/D_A$ varies but it does not exceed 6% in the best case as $D_B/D_A \to 0$. Hence, for small and medium $D_B/D_A$ values, the width offers a worse criterion than the shift between A and B profiles to discriminate between the concentrated and the dilute systems. However, for significantly large $D_B/D_A$ values, the relative difference of height vanishes whereas the relative difference of widths converges toward about 3%.

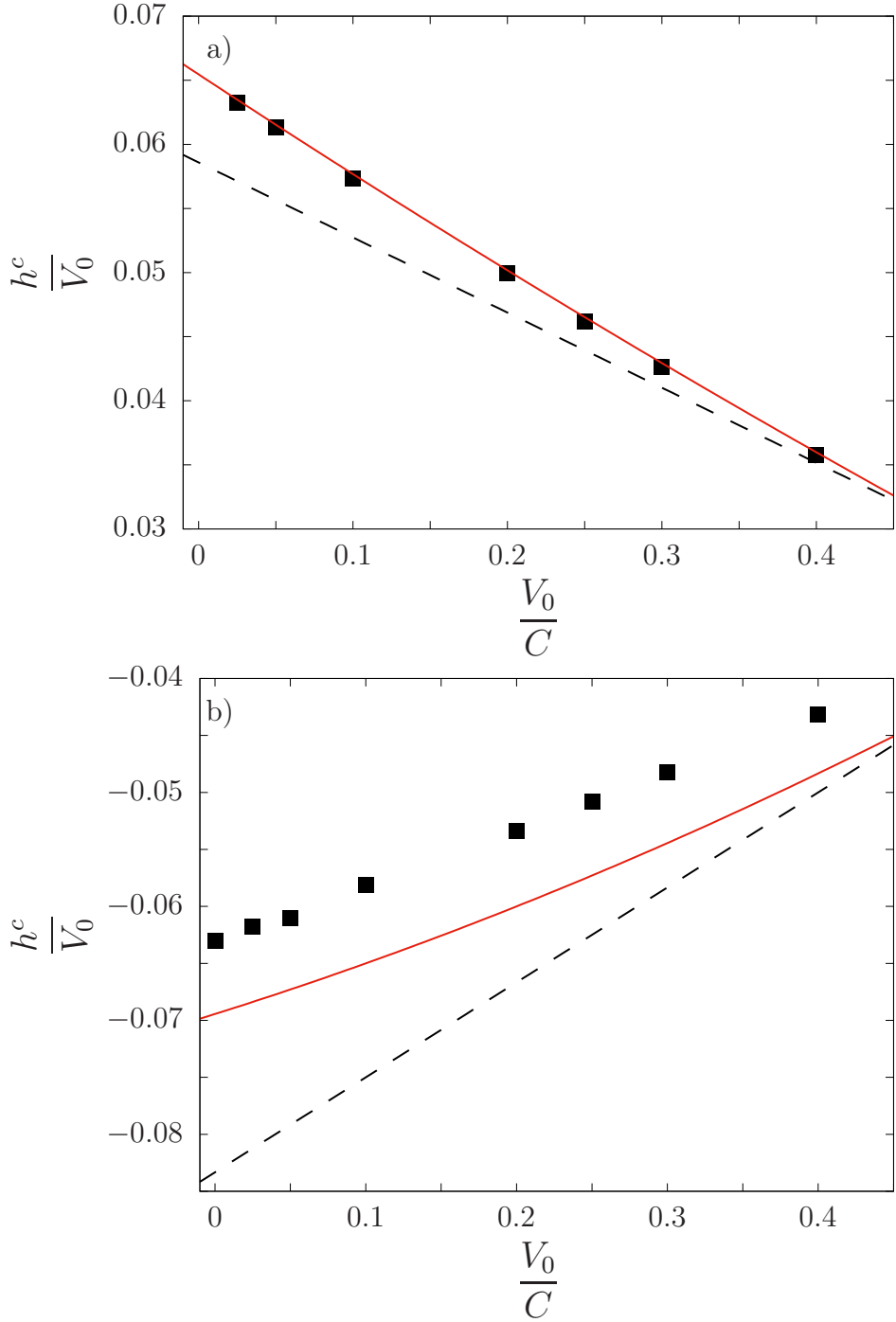

Figure 3: Scaled height $h^c/V_0$ versus concentration ratio $V_0/C$. First-order approximation (black dashed line), second-order approximation (red solid line), and numerical results (black squares) are presented for (a) $D_B/D_A = 1/16$ and (b) $D_B/D_A = 7/3$.

According to Eqs. (41) and (42), the front width $W^c$ in a concentrated system depends on $D_B/D_A$ and also the deviation from the dilution limit $V_0/C$. The variation of $W^c$ with $V_0/C$ is given in Fig. 5 for a sufficiently small ratio of diffusion coefficients $D_B/D_A = 1/16$ for the analytical prediction to be valid. As expected, the width $W^c$ tends to the value $W^d = 0.719$ associated with $D_B/D_A = 1/16$ for $V_0/C \to 0$. The width $W^c$ increases as $V_0/C$ increases: As the system becomes more concentrated, $W^c$ becomes closer to the width $W_0^d = W_0^c = 0.8$ obtained for $D_A = D_B$. Hence, the difference between the profile shape in a concentrated system with $D_B \neq D_A$ and the profile shape in a system with $D_A = D_B$ decreases as $V_0/C$ increases. We already came to an analogous conclusion for the variation of the height $h^c$ with $V_0/C$ as shown in Fig. 3. In conclusion, high concentrations tend to reduce the asymmetry of the profiles induced by the difference of diffusivities $D_A$ and $D_B$. This phenomenon is due to cross-diffusion in which the diffusion of both species influences the dynamics of each other.

# 5 Conclusion

In this work, we have studied the effects of concentration-induced perturbation of diffusion on FKPP wave fronts. The sensitivity of FKPP wave front to small perturbations makes it a good candidate for characterizing the effects of the deviation from the dilution limit on diffusion. We assume that high solute concentrations induce specific cross-diffusion terms in the reaction-diffusion equations in the framework of linear nonequilibrium thermodynamics. We consider two chemical species A and B engaged in the reaction $A + B \to 2A$ and with different diffusion coefficients, knowing that the perturbation of diffusion vanishes in the limit of identical diffusion coefficients. The analytical results deduced from a linear stability analysis show that the propagation speed in a concentrated system is not appreciably affected by the perturbation of diffusion. The relative correction of the profile width with respect to the dilute case presents an interesting behavior: It changes sign as the ratio of diffusion coefficients varies. However, the relative correction is in the order of 6% for concentration values in the domain of validity of the reaction-diffusion equations. We have introduced the height $h$ as the difference of concentrations between A and B species in the moving frame at the origin in order to evaluate the shift between A and B profiles due to their different diffusion coefficients. Contrary to the width, the relative correction to the height $h$ with respect to the dilute case reaches 28% for reasonable solute concentrations. The relative correction to the height is larger than 25% when the diffusion coefficient of species B is smaller than the one of species A. The diffusion coefficient of species B has to become at least 30 times larger than the diffusion coefficient of species A for the relative correction to the height to drop below 5%. In the limit of very large diffusion coefficients of species B, the width of the profile may be chosen as an alternative criterion to detect concentration-induced perturbations since the relative correction to the width converges to about 3% in this limit. We conclude that the FKPP wave front offers the opportunity to characterize concentration-induced perturbation of diffusion. Specifically, the shift of the concentration profiles of species associated with different diffusion coefficients is a well-adapted criterion showing significant variations with the deviation from the dilution limit in a wide range of diffusion coefficients.

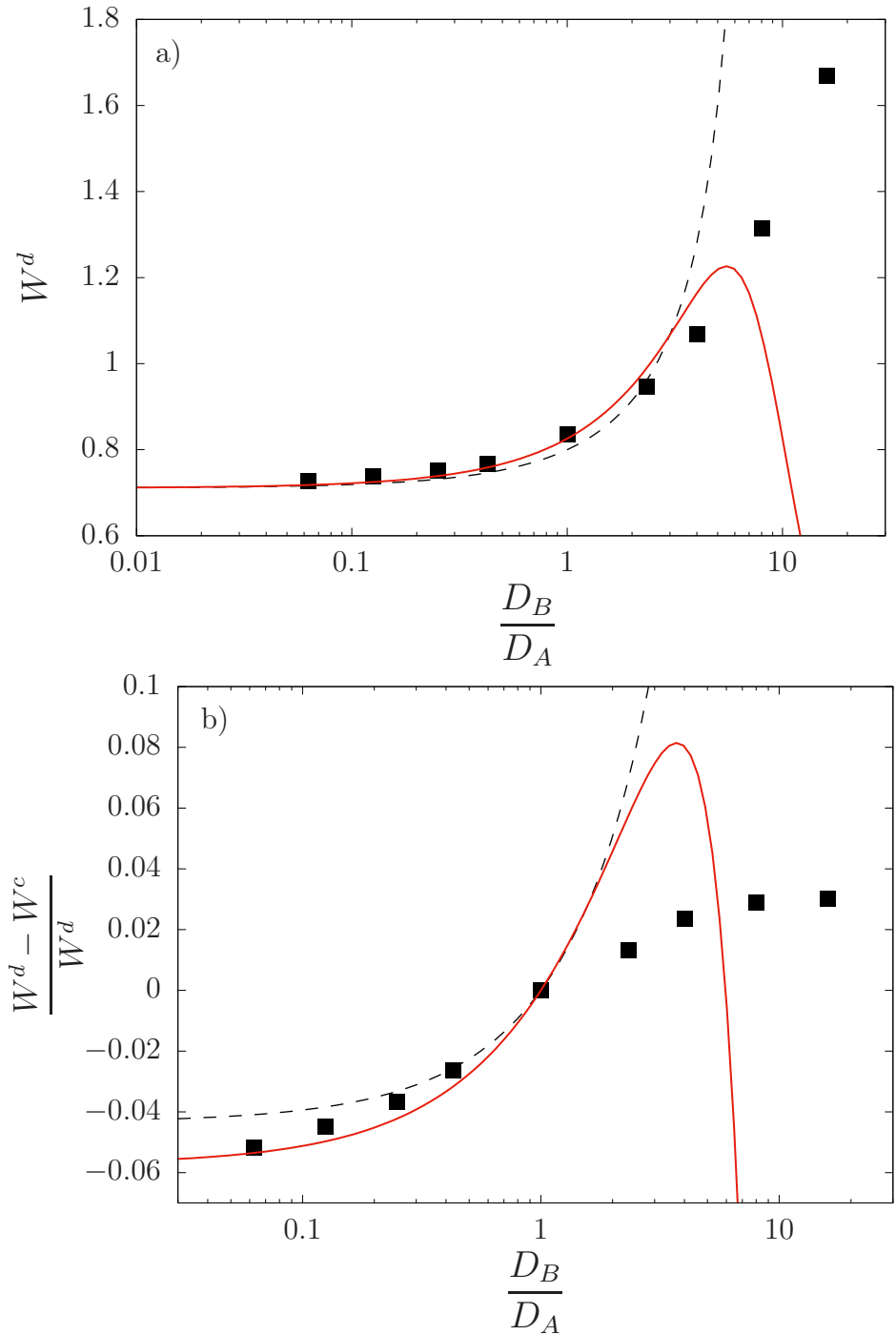

Figure 4: (a) Width $W^d$ in the dilution limit and (b) relative difference between the dilute case and the concentrated case for $V_0/C = 0.25$ versus the ratio $D_B/D_A$ of diffusion coefficients. First-order approximation (black dashed line), second-order approximation (red solid line), and numerical results (black squares).

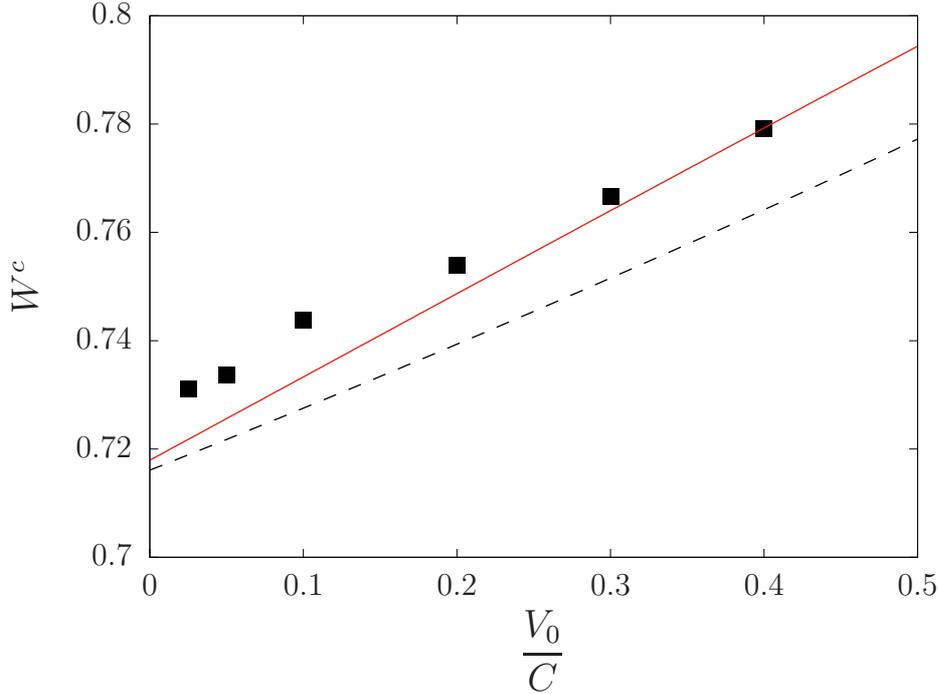

Figure 5: Width $W^c$ versus concentration ratio $V_0/C$. First-order approximation (black dashed line), second-order approximation (red solid line), and numerical results (black squares) are presented for $D_B/D_A = 1/16$.

# 6 Acknowledgements

This publication is part of a project that has received funding from the European Union's Horizon 2020 research and innovation programme under the Marie Skłodowska-Curie Grant Agreement No. 711859 and from the financial resources for science awarded for the implementation of an international cofinanced project.